\def\wh{wormhole }

\def\beq{\begin{equation}}
\def\eeq{\end{equation}}
\def\bea{\begin{eqnarray}}
\def\eea{\end{eqnarray}}

\documentstyle[11pt]{article}
\def\double{\baselineskip 24pt \lineskip 10pt}
\input epsf

\begin{document}

\begin{titlepage} 
\vspace*{-60pt} 
 
\vspace*{10pt} 
\begin{center} 
\LARGE 
{\bf Wormholes, Gamma Ray Bursts and the Amount of Negative Mass in the
Universe\footnote{This essay received an ``Honorable Mention'' from the
Gravity
Research Foundation, 1998 - Ed.}}\\ 
\vspace{.8cm} 
\normalsize 
\large{Diego F. Torres$^{1,2}$, Gustavo E. Romero$^{3, 4}$ and Luis A.
Anchordoqui$^2$} \\ 
\normalsize 

\vspace{.6 cm} 
{\em $^1$Astronomy Centre, University of Sussex, 
Falmer, Brighton BN1 9QJ, United Kingdom\\
${^2}$Departamento de F\'{\i}sica,  Universidad Nacional 
de La Plata, C.C. 67, 1900 La Plata, Argentina
${^3}$Instituto Astron\^{o}mico e Geof\'{\i}sico, USP, Av. M. Stefano 4200, 
CEP 04301-904, S\~ao Paulo, ~Brazil\\
${^4}$Instituto Argentino de Radioastronom\'{\i}a, C.C. 5, 1894 Villa Elisa, 
Argentina\\}
\end{center} 
 
\vspace{.4 cm} 
\begin{abstract} 
\noindent  In this essay,  we assume that negative mass objects can exist in the
extragalactic space and analyze   
the consequences of their microlensing on light from distant Active
Galactic Nuclei. We find that such events have
very similar features to some observed Gamma Ray Bursts  
and use recent 
satellite data to set an upper bound
to the amount of negative mass in the universe.

\end{abstract} 
 
\end{titlepage} 

\section{Introduction}

\begin{flushright}
{\sl  -I see nobody on the road, said Alice.\\
-I only wish I had such eyes, the King remarked in  a fretful tone.\\
To be able to see Nobody! And at that distance too!\\[.3cm]
{\rm Through the Looking Glass}\\
Lewis Carroll}
\end{flushright}

\double

The possibility of constructing high speed transports has fascinated
mankind along all history. And with the advent of the twentieth
century concepts and technologies, specially those of General Relativity, this
quest is more alive and thriving than ever before. But it is obvious to note 
that one way in which one
can travel faster is, instead of increasing the speed, to reduce the path. Let 
us
take a two dimensional space as an example. Is a straight line the
shortest path between two points? Certainly not: just fold the space until the
two points coincide and make a hole; the path is zero and this is, without 
doubt, the shortest
distance. This could be, of course, not allowed 
by most game rules, but it is the very concept of wormhole physics.
A subject that can be traced back up to 1916, a year after Einstein's 
remarkable
discovery that matter curves spacetime \cite{1916}. 

Wormholes are analytical solutions of the Einstein field equations, and 
basically
represent shortcuts in spacetime.  By now, wormholes are
theoretical entities and we can speculate but not rely on their possible
existence. A key point is that to keep wormholes open, preventing them 
for pinching off towards a singularity, one has to thread them with exotic 
matter.
This kind of matter violates the weak energy condition, and although quantum 
effects (of order $\hbar$) and scalar fields violate this condition, it is far
from clear whether macroscopic quantities of exotic matter can exist in the 
universe.
If exotic matter does exist, wormholes might have negative total mass
\cite{motho,VISSER-BOOK}. 

In this essay we shall assume that natural wormholes, or other form
of negative mass matter exist, 
and we shall study the extragalactic microlensing scenario
when light from distant Active Galactic Nuclei (AGNs) is affected by them. 
We shall find the unexpected conclusion that this kind of lensing very much
resembles the main features of some Gamma Ray Bursts (GRBs) and shall make
some highly testable predictions. Afterwards, we shall reconsider
our initial hypothesis and, using recent satellite data about GRBs,
we shall
obtain the first upper limit on the amount of 
negative mass in the universe.

\section{The model}
\subsection{Microlensing by a negative mass}

\indent Gravitational 
microlensing by negative masses was introduced by Cramer et al.
in 1995, considering that the lensing object, typically a wormhole, could be in 
the
galactic halo \cite{CRAMER}. 
In the application we wish to develop, the lens should be
isolated in extragalactic space. This makes the formulae slightly different, as 
for
instance in the computation of distances, but the concept remains. 
The Einstein radius of a negative mass is given by
$R_e= (4G|M|D)^{ 1/2}/c$, where apart from the usual meaning of the 
constants $c$ and $G$, $D$ represents an effective lens distance. 
This is a model-dependent parameter; 
its general expression is $D=D_{ol}
D_{ls}/D_{os}$, 
where $D_{ol}$, $D_{ls}$ and $D_{os}$ are the observer-lens, lens-source and 
observer-source angular diameter distances, all them computed as in 
\cite{NARAYAN}.
The variability timescale $T$ of a microlensing event is defined as the time 
that takes the line of sight to the source 
to cross the Einstein radius of the lens: $T=R_e/V$, while 
the overall relative intensity ${\rm I}_{{\rm neg}}$ is the modulation in 
brightness
of the background source detected by the observer. This is given by 
\cite{CRAMER},
\beq
{\rm I}_{{\rm neg}}=\frac { B^2 -2}{B \sqrt{B^2 -4} },
\eeq
where $ B(t)=B_0 ( 1 + ( t/t_v )^2 )^{1/2}$. Here, $B_0$ is a dimensionless 
impact parameter and $t_v$
is the transit time across the distance of the minimum impact parameter, $t_v
\propto T$. Taking ${\rm I}_{{\rm neg}}=0$ for $|B|<2$, it is possible
to obtain the light enhancement profiles for a negative amount of mass $M$.
These curves can be divided in 
two groups 
(see Fig. 3 of Ref. \cite{CRAMER}). 
For $B_0>2$, the light profiles are similar to the positive mass
cases but provide bigger light enhancement than that given by a similar 
amount of positive mass. For $B_0<2$, the curves are sharper 
and present 
divergences (caustics) of the intensity with an 
inmediate drop to zero. 
This happens at two times, solutions of $B^2-4=0$;
thus, for time running from $-\infty$ to $+\infty$, and during the same 
microlensing event, we obtain two divergences 
and two drops, of specular character.  
Unlike the $B_0>2$ case, these individual bursts 
present light profiles asymmetric under time reversal.

Although this also happens  in the usual (i.e. positive mass) scheme, 
two points should be noticed. First, the infinities 
arise from the geometrical approach based on point
mass objects. Any physical extent leads to 
finite amplifications \cite{SCHNEIDER-BOOK}. 
Second, a critical requirement for such a microlensing event to 
occur is that the size of 
the background source projected onto the lens plane must not be larger 
than the 
Einstein ring of the lensing mass \cite{chang}.
Background sources whose scale size is a fraction of the 
Einstein radius are amplified by significant factors, while the amplification
of sources whose 
projected sizes largely exceed the Einstein radius will be negligible.

\subsection{AGNs as background sources}

\indent AGNs are compact extragalactic sources of extraordinary 
luminosity that can radiate 
as much energy per unit of time as hundreds of normal galaxies. Recent satellite 
observations have shown that many, perhaps most, of these objects emit most of 
their power in the form of X- and $\gamma$-rays \cite{von}. The ultimate source 
of the energy of AGNs is widely believed to be accretion onto supermassive black 
holes. $\gamma$-ray emission is probably produced by inverse Compton scattering 
of ambient soft photons by ultrarelativistic electrons or positrons accelerated 
in the innermost part of the source \cite{BLANDFORD}. The energy spectrum of the 
resulting high energy radiation is a power law with indices roughly 
between 1.5 and 3. 
The emission regions have different size scales according to the 
energy range of the radiation due to $\gamma$-ray absorption by pair production 
in the radiation field of the central source. The inner regions are very 
compact ($\sim 10^{15}$ cm) and can constitute excellent 
background sources for 
microlensing. The idea that GRBs could be the 
result of microlensing events was proposed ten years ago by McBreen \& Metcalfe 
\cite{MCBREEN}. However, it was ruled out by the evidence provided by 
BATSE instrument 
about the basic asymmetry that most GRBs exhibit under time reversal. 
As we have shown above, 
microlensing by negative mass can produce asymmetric light curves quite 
naturally and the argument does not apply in such a case. 

\section{Output: qualitative analysis }

\indent GRBs have such a huge variety of features that they might 
hardly be accounted by a unique and comprehensive model. In fact, it has
been previously proposed that GRBs
should be explained in sets of similarity \cite{MITROFANOV}.  
With this idea in mind, we return to the
main statement of this work up to this stage: a wormhole-like lensing
upon light of a background AGN very much resembles some characteristics
of observed GRBs. To explore this in more detail, let us take each of 
the
two parts of a $B_0 < 2$ event separately. Two main features of a GRB are 
inmediately reproduced: they {\it burst} and they are distributed isotropically 
in 
the
sky, which without further biases is a natural extension of the model.
Furthermore, 
coincident features are that the observed distribution of energy peaks 
above 
50 keV, and that the spectrum at higher energies is given by a power law
with exponents in the approximate range (1.5, 2.5). The durations of
individual bursting events are widespread from 30{\rm ms} to 100{\rm s};
this can be particularly modelled for each burst, assuming different  
extragalactic
velocities and masses for the lenses. Most bursts are, in this kind of lensing,
asymmetric. But some symmetric could also take place when $B_0>2$ and the 
background source is particularly strong.
The observation of counterparts at other energy bands, as well as the absence of 
them,
is simply explained by comparing the pojected size of the emitting region of the 
source at each 
wavelength with the Einstein radius.  It is a matter of fact that spectra of 
most bursts have a cutoff at energies of {\rm GeV}. At these energies, the 
$\gamma$-spheres of the background AGN can be large enough  so as to exceed
the Einstein radius and prevent the occurrence of microlensing. The same process 
could operate for other emission regions. The different sizes of the different 
emitting parts of the AGN will be reflected in different variability timescales, 
in such a way that this could
stand for the differential durations of the
counterparts in a particular GRB event. 

Recently, some authors have expressed their perplexity by the fact that some 
GRBs, like 970828, do not present visible counterparts at all \cite{PALMER}. 
Notice that the smaller optical region in AGNs is typically $\sim10^{16}$ cm. 
Thus, it is quite possible that $\gamma$-spheres of $10^{14}-10^{15}$ cm be 
gravitationally magnified while the optical flux remains unaffected. Moreover, 
since the optical region is comparable to the outer $\gamma$-regions, the 
cutoff in the energy spectrum of events with optical counterparts should occur 
at higher energies (several tens of GeV) than in pure $\gamma$-ray events 
(where the cutoff should be at a few GeV).

In addition, microlensing by wormholes provide us with a highly testable
prediction. Every asymmetric burst generated by this mechanism 
should repeat, or should have 
been repeated in the past. Moreover, this repetition phenomena should occur
in opposite regimes: first with rising times shorter than the decays and then,
vice-versa. Unfortunately, this kind of phenomena cannot be directly detected
with the present satellite-borne $\gamma$-ray 
telescopes.  The errors  in 
position measurements are about $4^0$, 
and although individual bursts with temporal profiles in
both regimes have been observed, 
repetition studies must be of a statistical nature \cite{REPETITIONS}. These 
studies suggest that repeaters can occur within a range from
2\% to 7\% of the whole sample of more than 1300 bursts.
This is also consistent with our model, because the observed 
number of bursts with 
rising faster than the decay is larger than that 
in the oposite regime \cite{OPOSSITEREGIME}. 
Timescales for repeating bursts seem to be between 
some months and a few years, and it is clear that we can always select 
the mass of the lens to fulfill this requirement.

\section{A bound upon negative matter}

\indent Although we have retained this essay as a qualitative presentation,
we have made simulations of temporal profiles on BATSE data files. We have 
found that with an assumed extragalactic velocity of 1000 {\rm km s$^{-1}$}
and a source-lens configuration of redshifts $z_l=0.25$ and $z_s=2.5$,
bursts like BATSE \#257 and \#1089 can be very well fitted by substellar masses 
of
 $\sim -0.1 M_\odot$ and timescales of the order of years.  Now, we shall use 
this
knowledge to reconsider our initial hypothesis and estimate how much negative 
mass
could be in the universe.

The basic assumption here will be quite general: if compact objects of negative 
mass
already exist in the intergalactic space, 
they must produce GRB-like microlensing 
phenomena. Thus, we can use the number of actually observed GRBs to determine an 
upper limit to the spatial density of negative mass in compact, wormhole-like 
objects. If $|\rho|$ is the density of negative
mass, which for simplicity we take as constant, the optical depth 
to microlensing is given by \cite{Pa},

\beq
\tau= \frac{2 \pi}{3}  \frac{G\, D_{os}^2 \, |\rho|}{c^2}.
\eeq
The number of microlensing events observed in a lapse $\Delta T$ is
$N=2n \tau \Delta t/ \pi T$,
where $n$ is the total number of 
background AGNs and $T$ is the typical timescale for microlensing.
Then, using both previous formulae in favor of $|\rho|$, we get
\beq
\label{negrho}
|\rho|=\frac 34 \frac{T}{\Delta t}
\frac Nn \frac {c^2}{G} \frac {1}{D_{os}^2}.
\eeq

In (\ref{negrho}) we have quantites of two different kinds. 
Most of the magnitudes involved are related to observation. We have in this
group the number of background sources $n \simeq 10^9$ which comes from the
number of AGNs in the Hubble Deep Field,  and the observed
number of BATSE triggers, $N=1121$ during the first $\Delta t =3$ {\rm years}
of operation. The angular diameter distance
of the source is also fixed because the cosmological distribution of AGNs
seems to peak somewhere between $z_s=2$ and $z_s=3$, so we can adopt
an intermediate value of $z_s=2.5$. On the other hand, we have one
model-dependent magnitude, the variablity timescale of the problem, $T$.
As $T \simeq R_e /V$, we note that both the mass and  velocity of the
lens, are degrees of freedom of (\ref{negrho}). As we do want to have
an upper bound on $|\rho|$ we shall choose a conservative extragalactic
velocity of 1000 {\rm km s$^{-1}$}. Concerning the mass, we choose $-0.1
M_\odot$, which is suggested by the fits as a possible 
typical value. With these figures, 
we obtain, 
\beq
| \rho |\leq 2.03 \times 10^{-33} \,{\rm g\, cm ^{-3}}.
\eeq

The {\it less than} symbol is due to the fact that we do not
expect every BATSE trigger to be caused by a \wh lensing effect. We could
use the $\simeq 5 \%$ of possible repeating sources \cite{REPETITIONS}
as the number of observed events, and then
lower about two orders of magnitude in $|\rho|$. We note also that a greater
lens velocity, very likely in the extragalactic medium, could also
reduce the quoted number by
an order of magnitude.
We conclude then that the given value of $|\rho|$ must be considered 
as a 
large upper bound on the possible amount of negative mass in the universe. 
It is clear, consequently, that this amount is too small indeed to produce 
any significant cosmological consequences (remember that 
the mass contribution due to galaxies in the universe is 
$\sim 6 \times 10^{-31}$ g cm$^{-3}$ and the critical density is 
$\sim 1.9 \times 10^{-29}$ {\rm g cm$^{-3}$}).

\section{Final comment}

\indent Do wormholes really exist? Is a macroscopic amount of negative 
matter possible? We do not yet know the answer, but we
have shown that if they do populate 
the intergalactic space, they should produce observational signatures arising 
from 
gravitational microlensing of the light from distant AGNs. Since these 
signatures are similar to some GRBs, we used the available information about 
them to calculate an upper limit to the density of these exotic objects. The 
improvement of the quality of the observations will enable us to make important
conclusions about wormhole physics in the near future.
Either 
lower limits to the number of GRB repetitions will be established or some
concrete cases of repetitions will arise. In the first case, observation will 
point
towards ruling out the existence of 
any significant amount of negative mass in the universe. In the second case, a 
study
of the rising and decaying times could strongly support this existence. If 
finally 
we arrive at the conclusion that there is no natural negative mass wormhole
in the sky, basic research in the field  
still might render 
useful results beyond pure theoretical knowledge. After all, mankind 
did not wait to observe natural automobiles in order to build one.

\section*{Acknowledgments}

The research on which this work is based was partially supported by the 
Argentine agencies 
CONICET  (D.F.T. and G.E.R) and  ANPCT (PMT-PICT 0388) (G.E.R),
the Brazilian agency FAPESP (G.E.R), the British Council \& Fundaci\'on
Antorchas (D.F.T) and the FOMEC program (L.A.A.).
D.F.T. holds a Chevening Scholarship of the British Council. G.E.R. is a Member
of CONICET. We gratefully acknowledge James Lidsey for a critical reading of this 
essay.

\frenchspacing 


\end{document}